\documentclass{article}
\usepackage{spconf,amsmath,graphicx,balance}
\usepackage{booktabs, array} 
\usepackage{multirow} 
\usepackage{xcolor}
\usepackage{verbatim}

\usepackage{cite,url}
\usepackage{hyperref}
\hypersetup{
    colorlinks=false,
    linkcolor=purple,
    filecolor=magenta,      
    urlcolor=purple,
}
\def\pgftextcircled#1{
        \raisebox{.9pt}{\textcircled{\raisebox{-.9pt}{#1}}}}

\usepackage[pscoord]{eso-pic}
\newcommand{\placetextbox}[3]{
\setbox0=\hbox{#3}
\AddToShipoutPictureFG*{
\put(\LenToUnit{#1\paperwidth},\LenToUnit{#2\paperheight}){\vtop{{\null}\makebox[0pt][c]{#3}}}%
}%
}%

\title{Spoofing attack augmentation: can differently-trained attack models improve generalisation?}

\name{Wanying Ge$^1$, Xin Wang$^2$, Junichi Yamagishi$^2$, Massimiliano Todisco$^1$ and Nicholas Evans$^1$
\thanks{This work is supported by the TReSPAsS-ETN project funded by the European Union’s Horizon 2020 research and innovation programme under the Marie Skłodowska-Curie grant agreement No.\ 860813. It is also supported by JST CREST Grants JPMJCR18A6 and JPMJCR20D3, MEXT KAKENHI Grants 21K17775 and 21H04906.}}
\address{$^1$EURECOM, France~~~~~~~~ $^2$National Institute of Informatics, Japan}

\begin{document}

\maketitle
\begin{abstract}
    A reliable deepfake detector or spoofing countermeasure (CM) should be robust in the face of unpredictable spoofing attacks. To encourage the learning of more generaliseable artefacts, rather than those specific only to known attacks, CMs are usually exposed to a broad variety of different attacks during training. Even so, the performance of deep-learning-based CM solutions are known to vary, sometimes substantially, when they are retrained with different initialisations, hyper-parameters or training data partitions. We show in this paper that the potency of spoofing attacks, also deep-learning-based, can similarly vary according to training conditions, sometimes resulting in substantial degradations to detection performance. Nevertheless, while a RawNet2 CM model is vulnerable when only modest adjustments are made to the attack algorithm,  those based upon graph attention networks and self-supervised learning are reassuringly robust. The focus upon training data generated with different attack algorithms might not be sufficient on its own to ensure generaliability; some form of spoofing attack augmentation at the algorithm level can be complementary. 
\end{abstract}

\placetextbox{0.5}{0.08}{\fbox{\parbox{\dimexpr\textwidth-2\fboxsep-2\fboxrule\relax}{\footnotesize \centering \copyright  2024 IEEE. Personal use of this material is permitted. Permission from IEEE must be obtained for all other uses, in any current or future media, including reprinting/republishing this material for advertising or promotional purposes, creating new collective works, for resale or redistribution to servers or lists, or reuse of any copyrighted component of this work in other works.\\This work has been accepted to the IEEE International Conference on Acoustics, Speech and Signal Processing.}}}


\section{Introduction}
\label{sec:intro}
    Spoofing and deepfake detection solutions are typically trained using large databases which contain spoofed utterances generated with a rich variety of different spoofing attacks~\cite{wang2020asvspoof,liu2023asvspoof2021}. The goal is to encourage generalisation so that the learning of artefacts which characterise spoofs generated with known algorithms, e.g.\ text-to-speech (TTS) and voice conversions (VC) systems, will facilitate the reliable detection of new, different attacks encountered after deployment.  

    Today's state-of-the-art detection solutions perform with near-to-perfect performance in the case of known attacks, and even with low error rates for previously unseen attacks. Even so, their complex, deep neural architectures are sensitive to differences in initialisation, hyper-parameters or training data partitions, leading to occasionally-substantial variation in detection performance~\cite{wang2021comparative}. The vulnerability of deep neural networks to even-benign perturbations is well known. Usually, the perturbations are specially crafted adversarial attacks generated in order to manipulate the behaviour of a given, known model or 
    classifier~\cite{taylor2021sensitivity,claesen2015hyperparameter}.  They can also be used to degrade the performance of classifiers even in black-box setting in which the internal working of a model are unknown to the adversary~\cite{moosavi2017universal,wu2022adversarial}. 

    Through our efforts to improve upon reliability and our exploration of vulnerabilities, like others~\cite{liu2019Adversarial, gomezalanis2021adversarial, kawa2023defence}, we have found that spoofing/deepfake detection solutions are similarly vulnerable to adversarial attacks~\cite{panariello2023malafide}. As we present in this paper, we have also found that the same detection solutions are even vulnerable to spoofing attacks generated using deep models learned with even-subtle differences to the training conditions. This implies that an adversary can easily overcome a detection solution simply by retraining a text-to-speech or voice conversion algorithm, even the same algorithm used for the generation of spoof/deepfake detection training data; some detection systems see spoofs generated with the same, but differently trained algorithm as an entirely different attack altogether. 

    The remainder of the paper is organised as follows.  In Section~\ref{sec:attack_CM} we describe the VITS spoofing attack and the countermeasures used in this work.  We describe the experimental setup in Section~\ref{sec:setup} and in particular the different parameterisations of the VITS algorithm which we used to generate spoofed data. Our results are presented in Section~\ref{sec:results} and~\ref{sec:v1unseen}.

\section{Attack and countermeasures}
\label{sec:attack_CM}

    Our work was performed with the VITS algorithm for spoof / deepfake attack generation and with three detection / countermeasure solutions.  They are described below.

    \subsection{VITS as a spoofing attack}
    \label{sec:vits}
    
    VITS~\cite{kim2021vits} is a popular variational auto-encoder \cite{kingmaAutoencodingVariationalBayes2014} (VAE) based TTS model which converts a phoneme sequence into a speech waveform.  We chose VITS for this study only because the training procedure is especially efficient and is performed in an end-to-end manner without the need for the separate training of duration and acoustic models, or neural vocoder. Synthesized speech is of high quality, potentially because of the integration of adversarial training, normalizing flow \cite{rezende2015variational}, and stochastic duration modeling \cite{kim2021vits}. 

    The adoption of VITS also supports the generation of speech data with varied tempo, intonation, and other suprasegmental features, even when using the same model and phoneme input. This is achieved by changing the power of two random noises (i.e., the standard deviation of Gaussian noise). The first random noise is transformed by the flow-based model~\cite{durkanNeural2019a} into discrete numbers which represent the duration of the input phonemes. The second random noise is used when latent acoustic features are sampled through the reparameterization trick from the VAE posterior distributions~\cite{kingmaAutoencodingVariationalBayes2014}, conditioned on the input phonemes and generated duration. Waveforms are then generated from the latent acoustic features. By changing the power of the random noises, we can use the same VITS model to generate speech data of varying characteristics. 

    \subsection{Countermeasures}
    
    \noindent\textbf{AASIST}\footnote{\url{https://github.com/clovaai/aasist}}~\cite{jung2022aasist} is a state-of-the-art end-to-end (E2E) spoofing countermeasure solution based on graph attention networks~\cite{veličković2018graph}. A sinc-layer front-end~\cite{ravanelli2018sincnet} is used to extract feature representations directly from raw waveform inputs. The back-end graph attention layers are used to integrate both temporal and spectral representations. Finally a readout operation and a fully connected output layer are used to generate detection scores.

    \noindent\textbf{RawNet2}\footnote{\url{https://github.com/asvspoof-challenge/2021/tree/main/LA/Baseline-RawNet2}}~\cite{tak2021rawnet2} is an E2E model composed of a sinc-layer, six residual blocks, a recurrent layer with gated recurrent unit (GRU), and a fully connected output layer. The residual blocks extract frame-level deep feature representations given the output from the sinc-layer. The GRU layer then aggregates the frame-level representations into utterance-level representations which are then passed to the output layer to generate detection scores. 

    \noindent\textbf{Self-supervised leaning with AASIST (SSL-AASIST)}\footnote{\url{https://github.com/eurecom-asp/SSL\_Anti-spoofing}}~\cite{tak2022ssl} combines the AASIST back-end with front-end feature extraction using a pre-trained wav2vec2.0 model~\cite{babu2021xls}. Pre-training is performed using 437k hours of bona fide speech utterances sourced from five speech databases, covering 128 different languages and more than 60k speakers.
    SSL-based front-ends are known to be robust to additive noise and reverberation, in addition to other external influences~\cite{mohamedSelfSupervised2022}.

    \section{Experimental setup}
    \label{sec:setup}
    
    We describe the databases and protocols used in this work, together with details of the VITS training conditions and data generation procedures, in addition to implementation details and metrics.

    \subsection{Database}
    
    VITS models are trained using the VCTK database~\cite{yamagishi2019vctk}. Because the CM architectures and hyper-parameters (described below) were all designed and optimised using the ASVspoof 2019 database, we follow the same data pre-processing pipeline used for its generation. We downsampled the VCTK data to 16~kHz and applied high-pass filtering with a cut-in frequency of 80 Hz before training the VITS models. Speech data synthesized using VITS also has a sampling rate of 16~kHz. CMs were trained using the same set of VCTK bona fide but synthesized data generated using different VITS models. 

    \begin{table}[!t]
    \caption{VITS training and generation settings across different sets ( '-' indicates identical settings to V1).}
    \centering
    \resizebox{\columnwidth}{!}{%
        \renewcommand{\arraystretch}{1}
        \begin{tabular}{cccccc}
        \toprule
        & \multicolumn{3}{c}{\textbf{Training}} & \multicolumn{2}{c}{\textbf{Noise std. in generation}}   \\
        \cmidrule(lr){2-4} \cmidrule(lr){5-6}
        \textbf{Set ID} & Train set & \#. Mel chan. & Seed & For acoustic feat.  & For duration \\
        \hline
        V1 & set-1 & 80 & seed-1 & 0.667 & 0.8   \\
        V2 & - & 40 & - & - & -   \\
        V3 & set-2 & - & - & - & -   \\
        V4 & - & - & seed-2 & - & -   \\
        \hline
        V1.2 & \multicolumn{3}{c}{same VITS model as V1} & - & - \\
        V1.3 & \multicolumn{3}{c}{same VITS model as V1} & 0.1 & -   \\
        V1.4 & \multicolumn{3}{c}{same VITS model as V1} & - & 0.1   \\
        V1.5 & \multicolumn{3}{c}{same VITS model as V1} & 0.1 & 0.1   \\
        \bottomrule
        \end{tabular}
    }
    \label{tab:con-vits1234}
    \end{table}

    \begin{table*}[!t]
    \caption{CM performance in terms of the EER (\%) in different training and testing conditions.}
    \centering
    \resizebox{\textwidth}{!}{%
        \renewcommand{\arraystretch}{1}
        \setlength\tabcolsep{2pt}
                \newcolumntype{C}{>{\centering\arraybackslash}p{0.072\textwidth}}
        \begin{tabular}{cCCC|CCC|CCC|CCC}
        \toprule
        & \multicolumn{3}{c}{\textbf{Trained on V1}} & \multicolumn{3}{c}{\textbf{Trained on V2}} & \multicolumn{3}{c}{\textbf{Trained on V3}} & \multicolumn{3}{c}{\textbf{Trained on V4}}  \\
        \cmidrule(lr){2-4} \cmidrule(lr){5-7} \cmidrule(lr){8-10} \cmidrule(lr){11-13} 
        \textbf{Tested on} & AASIST & RawNet2 & SSL-AASIST & AASIST & RawNet2 & SSL-AASIST & AASIST & RawNet2 & SSL-AASIST & AASIST & RawNet2 & SSL-AASIST\\
        \hline
        V1 & 0 & 0 & 0 & 0 & 13.27 & 0.04 & 0.03 & 6.17 & 1.37 & 0.27 & 12.60 & 0.57 \\
        V2 & 0.50 & 6.27 & 0.07 & 0 & 0.03 & 0 & 0.67 & 8.70 & 0.47 & 0.67 & 11.23 & 0.13 \\
        V3 & 2.43 & 8.50 & 0.03 & 2.20 & 18.00 & 0.10 & 0 & 0 & 0 & 1.73 & 10.60 & 0.07 \\
        V4 & 1.20 & 7.93 & 0 & 0.57 & 15.93 & 0.07 & 0.13 & 5.87 & 0.13 & 0 & 0.13 & 0 \\
        \hline
        V1.2 & 0 & 0.67 & 0 & 0 & 13.03 & 0.30 & 0 & 5.47 & 1.40 & 0.23 & 12.47 & 0.60 \\
        V1.3 & 0 & 0.03 & 0 & 0 & 7.20 & 0.57 & 0 & 2.00 & 2.03 & 0.07 & 6.2 & 1.03 \\
        V1.4 & 0 & 0.93 & 0 & 0.03 & 12.63 & 0.33 & 0.03 & 7.27 & 1.80 & 0.33 & 15.03 & 1.07 \\
        V1.5 & 0 & 0.10 & 0 & 0 & 6.63 & 0.83 & 0.03 & 2.10 & 2.80 & 0.10 & 7.80 & 1.33 \\
        \hline
        Pooled & 0.77 & 3.73 & 0.01 & 0.50 & 11.49 & 0.37 & 0.16 & 5.03 & 1.50 & 0.57 & 10.11 & 0.63 \\
        \bottomrule
        \end{tabular}
    }
    \label{tab:results-vits}
    \end{table*}

    \subsection{VITS conditions}
    \label{sec:vits-conditions}

    As shown in Table~\ref{tab:con-vits1234}, we prepared four data sets (namely V1, V2, V3, and V4) by varying three different VITS model configuration parameters. For the V1 VITS model, we generated four additional sets -- V1.2, V1.3, V1.4, and V1.5 -- by varying two additional hyper-parameters. The training conditions vary according to differences in \pgftextcircled{1} the training data, \pgftextcircled{2} the number of Mel channels and \pgftextcircled{3} the random seeds for model parameter initialisation. Generation conditions vary according to the standard deviation of the noise that is used to generate \pgftextcircled{4} acoustic features and \pgftextcircled{5} duration, as described in Sec.~\ref{sec:vits}. 
    
    For \pgftextcircled{1}, we generate data using Mel-scaled spectrograms wih 80 bands as in the original work~\cite{kim2021vits} and using 40 bands as a contrastive condition. For \pgftextcircled{2}, we first select 3,000 utterances at random from the VCTK dataset for the bona fide partition of the CM training set. The remaining data is then split into two subsets (set-1 and set-2 in Table~\ref{tab:con-vits1234}). Among each subset, 20,185 audio samples are used for VITS training whereas the remaining 100 utterances are used for validation. For~\pgftextcircled{3}, we used two random seeds for VITS training (seed-1 and seed-2 in Table~\ref{tab:con-vits1234}). 
    For~\pgftextcircled{4} and~\pgftextcircled{5}, we use the V1 VITS model to generate sets V1.2 - V1.5 using the noise standard deviation values listed in to last two columns of Table~\ref{tab:con-vits1234}. Note that, although V1.2 is generated using the same VITS model and noise standard deviation value as V1, the generated data are different to those of V1 on account of the different noise values.
    
    We use V1, V2, V3, and V4 for CM training and testing. Furthermore, data sets V1.2 to V1.5 are also used for testing (but not training).
    
    \subsection{Implementation and metrics}
    
    All VITS networks\footnote{\url{https://github.com/jaywalnut310/vits}} are trained with an initial learning rate of $2\times 10^{-4}$ and with a scheduling factor of $0.999^{\frac{1}{8}}$ after each epoch. The batch size is set to 50. Training is performed using two NVIDIA GeForce RTX 3090 GPUs and is stopped after 300k steps. The CMs are trained using codes available online and with their default settings. We use the equal error rate (EER) as the performance metric to evaluate spoofing detection performance.

\section{Results}
\label{sec:results}

    Results are presented in Table~\ref{tab:results-vits}. It shows EER estimates for each CM for various matched and mismatched training (row 1) and testing sets (column 1). In the case of matched training and testing data (row 3, column 2-4; row 4, column 5-7; row 5, column 8-10; row 6, column 11-13), the EER for all three CMs is zero or near-to. This is not surprising given that the CMs are trained using spoofed utterances which contain the same artefacts as the test data (utterances generated with the same algorithm and configuration). 
    
    EERs are higher under mismatched conditions. Given that the set of bona fide utterances in each condition is identical, the cause of increased EERs relates to differences in spoofed utterances. These results show that the artefacts corresponding to VITS-generated utterances vary according to the data with which the algorithm is trained. In the following, we examine these findings in further detail for each CM.

    \textbf{AASIST} -- while EERs increase under mis-matched conditions, they remain reasonably low, with some still below 1\%. EERs remain low in the case of synthetic data generated using the same model, but with differences to the generation conditions (V1 \& V1.2 - V1.5);
    
    \textbf{RawNet2} -- EERs are generally substantially higher under mismatched conditions. However, EERs vary more substantially across different generation conditions. For example, results for the V2 training condition (column 6) show an EER of 13.27\% for the V1 test set, but 6.63\% for the V1.5 test set. Similar differences are observed for V3 and V4  training conditions. In contrast to AASIST, RawNet2 struggles in generalising to differences in the generation conditions.
    
    \textbf{SSL-AASIST} -- while EERs remain relatively low, SSL-AASIST is less robust than AASIST to variability in the generation conditions. For example, for the V2 training set, the EER of 0.04\% for the V1 test set increases by almost 20 times to an EER of 0.83\% for the V1.5 test set. While the EER remains low, this result is perhaps somewhat surprising given that the SSL front-end extracts high-level representations that typically generalise well to different conditions.

\section{Can diversity help robustness?}
\label{sec:v1unseen}

    \begin{table}[t]
    \caption{Performance in terms of the EER (\%) for CMs trained on combined sets V2-V4 and tested against unseen V1 and V1.2-V1.5 attacks.}
    \centering        \renewcommand{\arraystretch}{1}
        \begin{tabular}{cccc}
        \toprule
        & \multicolumn{3}{c}{\textbf{Trained on V2-4}}  \\
        \cmidrule(lr){2-4}
        \textbf{Tested on} & AASIST & RawNet2 & SSL-AASIST\\
        \hline
        V1 & 0 & 2.2 & 0 \\
        V2 & 0 & 2.93 & 0 \\
        V3 & 0 & 0.47 & 0  \\
        V4 & 0 & 1.37 & 0 \\
        \hline
        V1.2 & 0 & 1.9 & 0  \\
        V1.3 & 0 & 0.77 & 0.03  \\
        V1.4 & 0 & 2.83 & 0  \\
        V1.5 & 0 & 0.87 & 0.03  \\
        \hline
        Pooled & 0 & 1.79 & 0.01  \\
        \bottomrule
        \end{tabular}
    \label{tab:results-v234}
    \end{table}   

    Results presented above show that an adversary can improve the potential to bypass a spoofing countermeasure by making subtle changes to an algorithm used to generated spoofed utterances. This is particularly evident in the case of the RawNet2 CM, but also for the AASIST and SSL-AASIST CMs, albeit to a lesser extent. We are hence interested to determine whether robustness can be improved by training the CM with spoofed utterances generated by multiple, differently configured attack algorithms. We test this hypothesis by training a CM using spoofed utterances generated with V2, V3 and V4 configurations (fixed standard deviation) and by testing it with spoofed utterances generated with V1-V1.5 configurations. Results are shown in Table~\ref{tab:results-v234}.

    For AASIST and SSL-AASIST CMs, training on V2, V3 and V4 sets results in zero or near-to zero EERs for all V1-V1.5 sets. The comparison of results in Tables~\ref{tab:results-vits} and~\ref{tab:results-v234} show that, in this case, the use of attacks generated with multiple, differently configured attack algorithms is beneficial in terms of CM generalisation.

    The same approach is also beneficial in the case of the RawNet2 CM.  The pooled EER of 1.79\% in the case of training with V2, V3 and V4 sets is substantially lower than the pooled RawNet2 EERs in Table~\ref{tab:results-vits}. Still, the variation in the EER is high and the RawNet2 CM remains vulnerable to some attack configurations.  Clearly, the RawNet2 CM is inferior to  AASIST and SSL-AASIST alternatives.

\section{Conclusions}

    Our work presented in this paper shows that a CM trained using spoofed data generated with one attack configuration may be vulnerable to those generated with the same algorithm when it is differently configured. We also show that CM generalisation can be improved simply by training using spoofed utterances generated with multiple, differently configured attack algorithms. This is form of data augmentation which is known to be beneficial across a host of related problems, including speaker recognition and spoofing detection. While the usual approaches involve the introduction of additive and convolutive noise, which can improve CM reliability in the case of varying acoustic and channel conditions. In contrast, spoofing attack augmentation is beneficial in improving generalisation in the case of varying spoofing attacks involving modest changes to an attack algorithm which might otherwise cause substantial degradation to detection performance.

    Naturally, more thorough testing of current CM solutions with other attack algorithms is necessary. It will be of interest to determine whether variation in other attack algorithms is of the same concern as it is for the VITS algorithm and also whether similar spoofing attack augmentation strategies are equally beneficial in improving CM generalisation. Also of interest is whether spoofing attack augmentation is beneficial in terms of generalisation to entirely different attacks rather than differently configured attacks. Last, we are curious to determine whether spoofing attack augmentation can help in the continuous and growing fight against adversarial attacks.

\balance
\bibliographystyle{IEEEbib}
\bibliography{strings,refs}
\end{document}